\ifx\pdfoutput\undefined
\else
  \pdfoutput=1 
\fi

\documentclass[aps,prb,twocolumn,amsmath,amssymb]{revtex4-2}

\usepackage[T1]{fontenc}
\usepackage[utf8]{inputenc} 
\usepackage{lmodern}        
\usepackage{amsmath,amssymb,bm}
\usepackage{mathtools}      
\usepackage{dcolumn}        
\usepackage{physics}        
\usepackage{xcolor}         
\usepackage{graphicx}       
\DeclareGraphicsExtensions{.pdf} 

\usepackage{microtype}      
\usepackage{enumitem}       

\usepackage[colorlinks=true,allcolors=blue]{hyperref}

\newcommand{\trh}{\mathrm{tr}\,}
\newcommand{\deth}{\mathrm{det}_h\,}
\newcommand{\1}{\mathbb{1}} 

\newif\ifdraft
\draftfalse            
\ifdraft
  
  \newcommand{\cin}[1]{\textcolor{red}{#1}}
\else
  \newcommand{\cin}[1]{#1}    
\fi

\begin{document}

\title{
Gradient flow and Bogomolny bounds for quantum metric actions
}

\author{Takahiro Fukui}
\affiliation{Department of Physics, Ibaraki University, Mito 310-8512, Japan}

\date{\today}

\begin{abstract}
We formulate gradient flow dynamics generated by two natural actions of the quantum metric for an isolated set of Bloch bands.  Specializing to two spatial dimensions, we derive Bogomolny-type lower bounds that relate these actions to the Chern number and show that the bounds are saturated by (anti-)holomorphic projector configurations.  
Along the flows, the actions decrease monotonically while the Chern number is conserved, 
giving a constructive route to simplify models within a fixed topological phase 
toward canonical, low-complexity representatives.
\end{abstract}


\maketitle

\section{Introduction}
Geometric properties of Bloch bands \cite{Provost:1980aa}, encoded in the quantum metric and the Berry curvature,
control a wide variety of physical responses in crystalline systems.  Paradigmatic examples include the quantized Hall conductivity determined by the Brillouin zone integral of the Berry curvature \cite{Thouless:1982uq,kohmoto:85}.
Whereas Berry curvature has received broad attention due to its topological implications, the quantum metric has been  increasingly recognized as an equally fundamental ingredient.  

In band geometry, it provides intrinsic length scales \cite{Kolodrubetz:2017aa},
governs the superfluid weight \cite{Peotta:2015aa,PhysRevB.95.024515},
and constrains Wannier localization \cite{Marzari:1997aa,PhysRevLett.82.370}, among other roles.
It also plays a central role in nonlinear responses, including shift and injection currents in optics~\cite{PhysRevB.61.5337,doi:10.1126/sciadv.1501524,PhysRevX.10.041041,PhysRevLett.129.227401},
and in nonlinear Hall transport:
beyond the Berry curvature dipole mechanism \cite{PhysRevLett.115.216806}, 
an intrinsic quantum metric contribution has been identified theoretically \cite{PhysRevB.108.L201405} 
and observed experimentally \cite{doi:10.1126/science.adf1506}.
In many-body systems, the quantum weight has also attracted much interest 
\cite{PhysRevB.62.1666,PhysRevResearch.7.023158}.
\cin{Related higher-order quantum geometric tensors, constructed from higher-order
gauge-invariant cumulants, have also been discussed in Ref.~\cite{PhysRevA.108.032218}.}

Recent progress has clarified tight links between the quantum metric and topology in Chern insulators. 
In particular, Refs. \cite{LEDWITH2021168646,Ozawa:2021vs,PhysRevB.104.045104} cast 
band geometry in a K\"ahler framework 
and established sharp inequalities relating the quantum volume, 
i.e., the Brillouin zone area measured by the quantum metric, to the first Chern number. 
In two dimensions, the quantum volume saturates its lower bound when the Brillouin zone is endowed with the K\"ahler structure induced from the Grassmannian via the band projector map.

\cin{Complementary theoretical developments include projector-based analyses of geometric bounds,
holomorphicity conditions, and ideal-band geometries in Chern insulators
\cite{1zg9-qbd6,8ng1-bwf6,qscv-qxqt,qxbl-qd4f}.
These works focus on static inequalities, generalized Landau-level or ideal-band constructions,
and a systematic gauge-invariant projector calculus, but do not study explicit gradient flows.
In parallel, flows of band projectors that monotonically drive bands toward
``Wilson-loop-ideal'' geometries saturating Wilson-loop bounds on the quantum
metric have been proposed in Ref. \cite{yu2025wilsonloopidealbandsgeneralidealization}.
}

\cin{Building on these insights and on projector-based formulations of band geometry
\cite{PhysRevB.104.045104,qscv-qxqt}, 
we formulate an action principle for the quantum metric that unifies such results and, 
crucially, yields explicit gradient flows.}
These flows monotonically decrease our quantum metric actions while preserving the Chern number, 
and they drive generic models within a fixed topological sector 
toward bound-saturating (anti-)holomorphic projector configurations. 
This variational viewpoint offers both a conceptual bridge to K\"ahler geometry and a practical algorithm to obtain canonical, low-complexity representatives of topological bands.

\cin{Beyond these conceptual aspects, projector gradient flows may also serve as a practical tool.
Within a fixed topological sector, they can in principle drive band projectors toward
more geometry-optimized configurations with smoother or more uniform quantum geometry.
Such flows may assist in constructing model Hamiltonians closer to ideal Chern
bands or generalized Landau levels, which is relevant for stabilizing interaction-driven phases
such as fractional Chern insulators.  From a numerical perspective, they can be viewed as
topology-preserving ``preconditioning'' dynamics for variational optimizations involving band
projectors.}

This paper is organized as follows. Section \ref{s:actions} defines two SL$(2,\mathbb Z)$-covariant 
actions of the quantum metric in a projector-only, gauge-invariant formulation and fixes our notation. 
Section \ref{s:flows} derives the corresponding  gradient flow equations
and proves that the actions are monotonically nonincreasing while the Chern number is conserved. 
In Sec. \ref{s:bogomolny}, we establish \cin{ Bogomolny-type lower bounds \cite{MantonSutcliffe2004} } 
in two dimensions 
and identify the (anti-)holomorphic first order equations that saturate them. 
Section \ref{s:example} illustrates the flows using the Wilson-Dirac model, 
showing the approach to canonical low-complexity representatives within a fixed Chern sector. 
Section \ref{s:summary} summarizes our results and outlines open directions. 

\section{Actions for the quantum metric}\label{s:actions}

Our goal in this section is to discuss geometric features of Bloch states using the quantum metric as the sole input. 
To this end, we work directly with the projector $P(k)$ onto an isolated set of bands and with the reciprocal-space
 metric $h_{\mu\nu}$. We then introduce actions built from the quantum metric 
 that are invariant under reciprocal-basis relabeling, i.e., under SL$(2,\mathbb{Z})$. 
 These actions provide a variational framework that captures the intrinsic geometry of the band subspace itself, 
 rather than that of any particular tight-binding realization. 

In Sec.~\ref{s:QMactions} we define two natural choices: (i) a trace-type action 
reminiscent of the Grassmannian sigma model, and (ii) a determinant-type ``quantum-volume'' action. 
These functionals will later serve as Lyapunov functionals for gradient flows 
that monotonically reduce geometric complexity within a fixed topological sector. 
Before introducing the actions, Sec.~\ref{s:metric_h} summarizes our notation for reciprocal space, 
and Sec.~\ref{s:qgt} briefly reviews the quantum geometric tensor 
and its relation to both the quantum metric and the Berry curvature.

\subsection{Conventions}\label{s:metric_h}

Let $k=k^\mu \bm b_\mu$ be the crystal momentum, where $\bm b_\mu$ are primitive reciprocal lattice vectors.
Greek indices $\mu,\nu=1,2$ label components in a  reciprocal basis, whereas Latin indices $a,b=1,2$ are 
reserved for an orthonormal frame specified later in Sec. \ref{s:bogomolny}.
Introduce the dual basis $\bm b^\mu$ defined by
\begin{alignat}{1}
\bm b^\mu\!\cdot\!\bm b_\nu=\delta^\mu_{\ \nu}.
\end{alignat}
Although $\bm b^\mu$ can be identified with real-space primitive vectors up to normalization, 
we will use them here purely as the dual basis in reciprocal space.

The reciprocal-space metric is
\begin{alignat}{1}
ds^2=h_{\mu\nu}\,dk^\mu dk^\nu,\qquad h_{\mu\nu}=\bm b_\mu\!\cdot\!\bm b_\nu,
\end{alignat}
with $h^{\mu\nu}=(h^{-1})^{\mu\nu}$ and $h=\det h_{\mu\nu}$.
The integration measure is
\begin{alignat}{1}
d\mu=\frac{1}{2}\,\epsilon_{\mu\nu}\,dk^\mu\wedge dk^\nu,
\end{alignat}
where the Levi-Civita tensors are defined by
\begin{alignat}{1}
\epsilon_{\mu\nu}=\sqrt{h}\,\varepsilon_{\mu\nu},\qquad
\epsilon^{\mu\nu}=\frac{1}{\sqrt{h}}\,\varepsilon^{\mu\nu},
\end{alignat}
and $\varepsilon_{12}\equiv 1$ denotes the totally antisymmetric symbol.

\subsection{Quantum geometric tensor}\label{s:qgt}

Define the derivatives $\partial_\mu=\partial/\partial k^\mu$ and let $P(k)$ be the Hermitian projector onto an isolated set of $N$ bands, so that $P^2=P=P^\dagger$.
For a generic multi-band projector, the quantum geometric tensor is defined by
\begin{alignat}{1}
Q_{\mu\nu}(k)=\Tr\, P\,\partial_\mu P\,\partial_\nu P
= g_{\mu\nu}(k) + \frac{1}{2}F_{\mu\nu}(k),
\end{alignat}
where the symmetric (real) and antisymmetric (purely imaginary) parts are
\begin{alignat}{1}
g_{\mu\nu}(k)&=\Re\,Q_{\mu\nu}(k)=\frac{1}{2}\Tr \partial_\mu P\,\partial_\nu P,\\
F_{\mu\nu}(k)&=2i\Im\,Q_{\mu\nu}(k)=\Tr P[\partial_\mu P,\partial_\nu P].
\label{GandF}
\end{alignat}
Note that with this convention $F_{\mu\nu}$ is purely imaginary.

Both $Q_{\mu\nu}$ and $g_{\mu\nu}$ are positive semidefinite in the sense of Gram matrices.
Define the matrix inner product $\langle X,Y\rangle=\Tr X^\dagger Y$,  and  introduce
$X_\mu=(\partial_\mu P)P$ and $Y_\mu=\partial_\mu P$.
Then, we have
$Q_{\mu\nu}=\langle X_\mu,X_\nu\rangle$ 
 and  $2g_{\mu\nu}=\langle Y_\mu,Y_\nu\rangle$,
so $Q_{\mu\nu}$ and $g_{\mu\nu}$ are Gram matrices and hence,  positive semidefinite.
It also turns out that $g$ is the real part of a hermitian positive semidefinite matrix $Q$, ensuring  
the positive semidefinitness of $g$.

\subsection{Actions of the quantum metric}\label{s:QMactions}

Motivated by the positive semidefiniteness of the quantum metric, 
we introduce two complementary actions that probe the geometry of a band subspace.

Trace-type action: 
A natural additive measure is the trace of the metric over momentum indices,
\begin{alignat}{1}
S_{\rm t}=\frac{1}{2}\int \trh g(k)\, d\mu,
\label{TrAct}
\end{alignat}
where $\tr g \equiv g^\mu{}_\mu=h^{\mu\nu} g_{\mu\nu}$ is the trace with respect to the reciprocal-space metric $h_{\mu\nu}$. 
This functional is closely related to Wannier localization measures \cite{Marzari:1997aa,PhysRevLett.82.370}.
In particular, in an orthonormal frame, Eq. (\ref{TrAct}) 
coincides, up to the conventional Brillouin zone normalization,  
with the gauge-invariant part of the Marzari-Vanderbilt spread functional
for an isolated composite band \cite{Marzari:1997aa}.

Determinant-type action: 
A multiplicative notion of local geometric size is provided by the determinant,
\begin{alignat}{1}
S_{\rm d}=\int \sqrt{\deth g(k)}\, d\mu,
\end{alignat}
with $\deth g \equiv \det(h^{-1}g)=\det g/\det h$, so that in two dimensions $\sqrt{\deth g}\, d\mu=\sqrt{\det g}\, dk^1\wedge dk^2$.
The integrand $\sqrt{\det g}$ measures the Brillouin zone area induced by $g$, 
whose sharp lower bounds and K\"ahler-saturation properties were analyzed in 
Refs. \cite{LEDWITH2021168646,Ozawa:2021vs,PhysRevB.104.045104}. 
Here we recast that quantity as a variational action to be used as a Lyapunov functional for gradient flows.

\cin{Intuitively, $S_{\rm t}$ measures the average local magnitude of the quantum metric:
large values signal rapidly varying projectors and strongly curved band geometry, whereas
decreasing $S_{\rm t}$ smooths the Bloch states in momentum space.  In contrast, $S_{\rm d}$
measures the total quantum volume of the Brillouin zone induced by $g_{\mu\nu}$; minimizing
$S_{\rm d}$ drives the metric toward more isotropic, K\"ahler-like configurations compatible
with the topological constraints.}

By construction, both $S_{\rm t}$ and $S_{\rm d}$ are invariant under reciprocal-basis relabeling 
$\bm b'_\mu = M_\mu{}^\nu \bm b_\nu$ with $M\in\mathrm{SL}(2,\mathbb Z)$.
\cin{Here SL$(2,\mathbb Z)$ corresponds to integer changes of basis in reciprocal space,
which generate different fundamental parallelograms of the same Brillouin-zone torus while
preserving its area and orientation.  Equivalently, they relabel the same reciprocal lattice
by choosing different primitive reciprocal vectors, so that the components of $h_{\mu\nu}$ and
$g_{\mu\nu}$ change even though the underlying torus and band geometry do not.  Covariance under
SL$(2,\mathbb Z)$ therefore guarantees that our actions depend only on the Brillouin-zone torus
and the projector $P(k)$, not on a particular choice of primitive reciprocal vectors.}

\section{Gradient flow equations}\label{s:flows}

We now set up the variational calculus needed to derive the flows generated by the quantum metric actions. 
Our basic field is the Hermitian projector $P(k)$ onto an isolated set of bands, 
so all admissible variations must preserve the constraint $P^2=P$ to first order. 
Differentiating $P^2=P$ gives
$\delta P^2=\delta P\,P+P\,\delta P=\delta P$,
which implies that the diagonal blocks of $\delta P$ vanish:
\begin{alignat}{1}
P\,\delta P\,P=(1-P)\,\delta P\,(1-P)=0.
\end{alignat}
Equivalently, any admissible Hermitian variation can be parametrized in off-diagonal form as
\begin{alignat}{1}
\delta P=(1-P)X P + P X^\dagger(1-P),
\label{Var}
\end{alignat}
with an arbitrary matrix $X$. In what follows we use \eqref{Var} to project the forces generated 
by the two actions onto the tangent space of the projector manifold 
and thereby obtain the corresponding gradient flow equations.

\cin{From a physical perspective, the resulting dynamics may be viewed as
``geometry-simplifying'' flows for band projectors: they monotonically reduce a chosen
quantum-metric action while preserving the Chern number, and hence could, in principle, drive $P(k,\tau)$
toward smoother or more holomorphic geometries within a fixed topological sector.}

\subsection{Trace action}

We begin with the trace-type functional on the Brillouin zone torus, 
where integration by parts produces no boundary terms. 
Variations must preserve the projector constraint $P^2=P$ and are 
therefore off-diagonal in the sense of Eq.~\eqref{Var}. The first variation reads
\begin{alignat}{1}
 \delta S_{\rm t}
 = -\frac{1}{2}\int \Tr\,(\Delta P)\,\delta P\, d\mu,
 \label{VarTr}
\end{alignat}
where $\Delta\equiv \partial_\mu\partial^\mu$ is the Laplacian built from the background metric 
$h_{\mu\nu}$ (with $\partial^\mu=h^{\mu\nu}\partial_\nu$). 
A naive descent by $\Delta P$ would not, in general, preserve $P^2=P$; 
one must project the force onto the tangent space of the projector manifold by enforcing the off-diagonal form \eqref{Var}.
Substituting Eq.~\eqref{Var} into Eq.~\eqref{VarTr}, we obtain
\begin{alignat}{1}
\delta S_{\rm t}
 = -\frac{1}{2}\int \Tr\big[P (\Delta P)&(1-P)X
 +(1-P)(\Delta P)P\, X^\dagger\big] d\mu .
\end{alignat}
Using
$PA(1-P)=P[P,A](1-P)$ and $(1-P)AP=-(1-P)[P,A]P$ for any matrix $A$ and  
applying these commutator identities twice yields the projected form
\begin{alignat}1
 \delta S_{\rm t}
 =& -\frac{1}{2}\int \Tr\!\big(P[P,\Delta P](1-P)X
 \nonumber\\
 &\qquad-(1-P)[P,\Delta P]PX^\dagger\big)\,d\mu 
 \nonumber\\
 =& -\frac{1}{2}\int \Tr\!\big(P[P,[P,\Delta P]](1-P)X
 \nonumber\\
 &\qquad+(1-P)[P,[P,\Delta P]]PX^\dagger\big)\,d\mu 
 \nonumber\\
 =& -\frac{1}{2}\int \Tr\,[P,[P,\Delta P]]\,\delta P\,d\mu .
 \end{alignat}
Hence the gradient flow that both descends $S_{\rm t}$ and preserves $P^2=P$ 
takes the double commutator form
\begin{alignat}{1}
\partial_\tau P(k,\tau)
= -\frac{\delta S_{\rm t}}{\delta P(k,\tau)}
= \frac{1}{2}\,[P,[P,\Delta P]],
\label{TraGraFlo}
\end{alignat}
which is manifestly tangent to the projector manifold and therefore keeps $P^2=P$ for all~$\tau$.

\subsection{Determinant action}

We now turn to the determinant-type functional, which weights the projector gradients by the local quantum volume density $\sqrt{\deth g}$. Varying the density gives
\begin{alignat}{1}
\delta \sqrt{\deth g}
&=\frac{1}{2}\sqrt{\deth g} \,(g^{-1})^{\mu\nu}\delta g_{\nu\mu},
\end{alignat}
where \((g^{-1})^{\mu\nu}\) denotes the matrix inverse of \(g_{\mu\nu}\) 
written with raised indices via the background metric \(h\), more precisely
$(g^{-1})^{\mu\nu} \equiv \big((h^{-1}g)^{-1}h^{-1}\big)^{\mu\nu}$.
Using this, the first variation of the action reads
\begin{alignat}{1}
\delta S_{\rm d}=-\frac{1}{2}\int\Tr\,\partial_\mu\!\big(\sqrt{\deth g}(g^{-1})^{\mu\nu}\partial_\nu P\big)\,\delta P\,d\mu.
\end{alignat}
As in the trace case, the force must be projected onto the tangent space of the projector manifold. 
Imposing the off-diagonal variation (\ref{Var}) gives
\begin{alignat}{1}
\delta S_{\rm d}=-\frac{1}{2}\int\Tr\!\big[P,\,[P,\,\partial_\mu(\sqrt{\deth g}(g^{-1})^{\mu\nu}\partial_\nu P)]\big]\delta P\,d\mu.
\end{alignat}
Therefore the projector-preserving steepest descent takes the double commutator form, as in the case of the 
trace-action (\ref{TraGraFlo}), 
\begin{alignat}{1}
\partial_\tau P=\frac{1}{2}[P,[P,\partial_\mu(\sqrt{\deth g}(g^{-1})^{\mu\nu}\partial_\nu P)]].
\label{DetGraFlo}
\end{alignat}
This flow is gauge invariant, SL$(2,\mathbb Z)$-covariant, and manifestly tangent to the projector manifold, so it preserves $P^2=P$ for all~$\tau$, as in the case of the trace-action.

\subsection{Actions decreasing under gradient flow}\label{s:actdec}

We have derived the gradient flow equations \eqref{TraGraFlo} and \eqref{DetGraFlo} for the two action functionals. 
Here we show that, along either flow, the corresponding action is monotonically nonincreasing.

Both flows can be written in the unified form
\begin{alignat}{1}
\partial_\tau P=[P,J],\qquad J=[P,K],
\end{alignat}
up to a harmless rescaling of the flow time $\tau$, absorbing the prefactor $1/2$. 
For the trace action one may take $K=\Delta P$, and for the determinant action
$K=\partial_\mu\!\big(\sqrt{\deth g}\,(g^{-1})^{\mu\nu}\partial_\nu P\big)$.
In either case,  $P=P^\dagger$, $K=K^\dagger$, and hence $J^\dagger=-J$.

Along both flows the actions decrease according to
\begin{alignat}{1}
\partial_\tau S&=-\frac{1}{2}\int \Tr K\partial_\tau P \,d\mu
=-\frac{1}{2}\int \Tr K[P,J]\,d\mu
\nonumber\\
&=-\frac{1}{2}\int \Tr [K,P]J\,d\mu
=-\frac{1}{2}\int\Tr J^\dagger J\,d\mu\le0.
\end{alignat}
Thus each action is nonincreasing along its gradient flow. Equality holds iff $J=0$, i.e., at stationary points of the flow, and the actions are then expected to approach the Bogomolny lower bounds discussed next.

\section{Bogomolny bound}\label{s:bogomolny}

The actions introduced above admit \cin{Bogomolny-type \cite{MantonSutcliffe2004}} lower bounds 
\cite{PhysRevB.90.165139,LEDWITH2021168646,Ozawa:2021vs,PhysRevB.104.045104}. 
To derive them, 
it is convenient to pass to an orthonormal frame and subsequently to complex coordinates on the Brillouin zone. 
We introduce a vielbein $e^a{}_\mu$ such that
\begin{alignat}{1}
 h_{\mu\nu}=\delta_{ab}e^a{}_\mu e^{b}{}_\nu,\quad
 \delta^{ab}=h^{\mu\nu}e^a{}_\mu e^b{}_\nu,
 \end{alignat}
which realizes the background metric $h_{\mu\nu}$ as the pullback of the Euclidean metric $\delta_{ab}$. 
Multiplying the first relation by $h^{\nu\lambda}$ yields
\begin{alignat}{1}
\delta_\mu^\lambda =h_{\mu\nu}h^{\nu\lambda}=
\delta_{ab} e^a{}_\mu e^{b}{}_\nu h^{\nu\lambda}
=e^a{}_\mu e_a{}^\lambda.
 \end{alignat}
Thus $e^a{}_\mu$ and $e_a{}^\mu$ are mutual inverses, summarized by
$
 e_a{}^\mu e^a{}_\nu  =\delta^\mu_\nu$ and 
 $e_a{}^\mu e^b{}_\mu=\delta_a^b
 $.
In this orthonormal frame the Levi-Civita tensors become the simple totally antisymmetric symbol:
\begin{alignat}{1}
\epsilon_{ab}=e_a{}^\mu e_b{}^\nu \epsilon_{\mu\nu}=\varepsilon_{ab},
\qquad
\epsilon^{ab}=\varepsilon^{ab},
\end{alignat}
so volume forms and tensor contractions reduce to their flat expressions. We also define frame derivatives
\begin{alignat}{1}
\partial_a=e_a{}^\mu\partial_\mu,
\end{alignat}
and introduce the complex combinations
\begin{alignat}{1}
\partial=\partial_{a=1}-i\partial_{a=2},\qquad
\bar\partial=\partial_{a=1}+i\partial_{a=2}.
\end{alignat}
With these preparations in place, we now derive the Bogomolny bounds for the trace and determinant actions in turn.

\subsection{Trace action}

In the orthonormal frame the trace functional reads
\begin{alignat}{1}
S_{\rm t}=\frac{1}{4}\int \Tr \partial^a P\partial_a P\, d\mu,
\end{alignat}
with $d\mu\equiv \tfrac{1}{2}\epsilon_{ab}\,dk^a\wedge dk^b=dk^{a=1}\wedge dk^{a=2}$ in this frame. The key algebraic identity is
\begin{alignat}{1}
\partial^aP\,\partial_aP
=\left\{\begin{array}{c}\partial P\,\bar\partial P \\[2pt] \bar\partial P\,\partial P\end{array}\right\}
\mp i\,\epsilon^{ab}\,\partial_a P\,\partial_b P,
\end{alignat}
which splits the kinetic term into a manifestly positive part and a term proportional to the antisymmetric tensor. 
In addition, one has
\begin{alignat}{1}
\Tr \partial^aP\,\partial_aP=2\,\Tr \,P\,\partial^aP\,\partial_aP,
\end{alignat}
as follows from
\begin{alignat}{1}
\Tr P\,\partial^aP\,\partial_aP
&=\Tr \partial^aP\,(1-P)\,\partial_aP
\nonumber\\
&=\Tr (1-P)\,\partial_aP\,\partial^aP.
\end{alignat}
Putting these together, we can complete the square:
\begin{widetext}
\begin{alignat}{1}
S_{\rm t}
&=\frac{1}{2}\int \Tr P\,\partial^aP\,\partial_a P\,d\mu
=\frac{1}{2}\int \Tr\left\{
\begin{array}{c}
(\bar\partial P\,P)^\dagger(\bar\partial P\,P)\\[2pt]
(\partial P\,P)^\dagger (\partial P\,P)
\end{array}
\right\}\,d\mu
\mp \frac{i}{2}\int \epsilon^{ab}\,\Tr P\,\partial_a P\,\partial_b P \,d\mu
\nonumber\\
&\ge
\mp\frac{i}{2}\int \varepsilon^{\mu\nu}\,\Tr P\,\partial_\mu P\,\partial_\nu P \,d^2k
=\mp\pi\, C,
\label{TrBogBou}
\end{alignat}
\end{widetext}
where we used $\epsilon^{ab}\,\partial_aP\,\partial_bP=\epsilon^{\mu\nu}\,\partial_\mu P\,\partial_\nu P
=\varepsilon^{\mu\nu}\,\partial_\mu P\,\partial_\nu P/\sqrt{h}$. 
Here and below,
$d^2k \equiv dk^{\mu=1}\wedge dk^{\mu=2}$
denotes the coordinate two-form in the $(\mu,\nu)$ basis, so that
$d\mu=\tfrac{1}{2}\epsilon_{\mu\nu}\,dk^\mu\wedge dk^\nu
=\sqrt{h}\,d^2k$
in general.
$C$ denotes the Chern number
defined by
\begin{alignat}{1}
C=\frac{i}{2\pi}\int \varepsilon^{\mu\nu}\,\Tr P\,\partial_\mu P\,\partial_\nu P\,d^2k
=\frac{i}{2\pi}\int F_{12}\,d^2k,
\label{CheNum}
\end{alignat}
where $F_{12}$ is defined in Eq. (\ref{GandF}).
Therefore, from Eq. (\ref{TrBogBou}) we obtain the Bogomolny bound  \cite{Ozawa:2021vs,PhysRevB.104.045104}
\begin{alignat}{1}
S_{\rm t}\ge \pi\,|C|.
\label{TrBog}
\end{alignat}
The bound is saturated precisely when the positive term vanishes, i.e., when $P$ satisfies the first-order equations
\begin{alignat}{1}
\left\{\begin{array}{ll}
 \bar\partial P\,P=0 & \quad (C<0),\\[2pt]
 \partial P\,P=0 & \quad (C>0),
\end{array}\right.
\label{Hol}
\end{alignat}
namely, (anti-)holomorphic projector configurations in the orthonormal frame
 \cite{Ozawa:2021vs,PhysRevB.104.045104}.

\subsection{Determinant action}

In the orthonormal frame the determinant-type functional reads
\begin{alignat}{1}
S_{\rm d}=\int \sqrt{{\det}_\delta g}\,d\mu \;=\; \int \sqrt{\det(g_{ab})}\,d\mu.
\end{alignat}
The lower bound follows from the positive semi-definiteness of the quantum geometric tensor, 
as discussed in Sec. \ref{s:qgt}. 
In the present frame,
\begin{alignat}{1}
Q_{ab}=g_{ab}+\frac{1}{2}F_{ab}
\end{alignat}
is positive semi–definite, hence
\begin{alignat}{1}
\det (Q_{ab})=\det (g_{ab})-\Big(\frac{iF_{12}}{2}\Big)^2\;\ge\;0,
\end{alignat}
with equality iff $X_{a=1}=i\lambda(k)\,X_{a=2}$ for some function $\lambda(k)$, where $X_a=(\partial_a P)P$.
Therefore,
\begin{alignat}{1}
S_{\rm d}=\int \sqrt{\det (g_{ab})}\,d\mu
\;\ge\; \pi\left|\frac{i}{2\pi}\int F_{12}\,d\mu\right|.
\end{alignat}
As in the trace-action case, using $F_{12}\,d\mu\equiv\epsilon^{ab}\Tr P\partial_aP\partial_bP\,d\mu
=\epsilon^{\mu\nu}\Tr P\partial_\mu P\partial_\nu P \,d\mu
=\varepsilon^{\mu\nu}\Tr\!\big(P\partial_\mu P\partial_\nu P\big)\,d^2k$,
the right-hand-side becomes $\pi|C|$, and we obtain
\begin{alignat}{1}
S_{\rm d}\;\ge\; \pi|C|.
\label{DetBog}
\end{alignat}
Saturation occurs iff the projector satisfies the first-order weighted (anti-)holomorphicity condition,
\begin{alignat}{1}
(\partial_1P - i\,\lambda(k)\,\partial_2P)P=0.
\label{CauRie}
\end{alignat}
In the special cases $\lambda=\pm 1$, this reduces to Eq.~(\ref{Hol}).
These first-order equations at saturation were discussed in detail in Refs.~\onlinecite{Ozawa:2021vs,PhysRevB.104.045104}.

\begin{figure*}[t]
\centering
\includegraphics[width=0.25\linewidth]{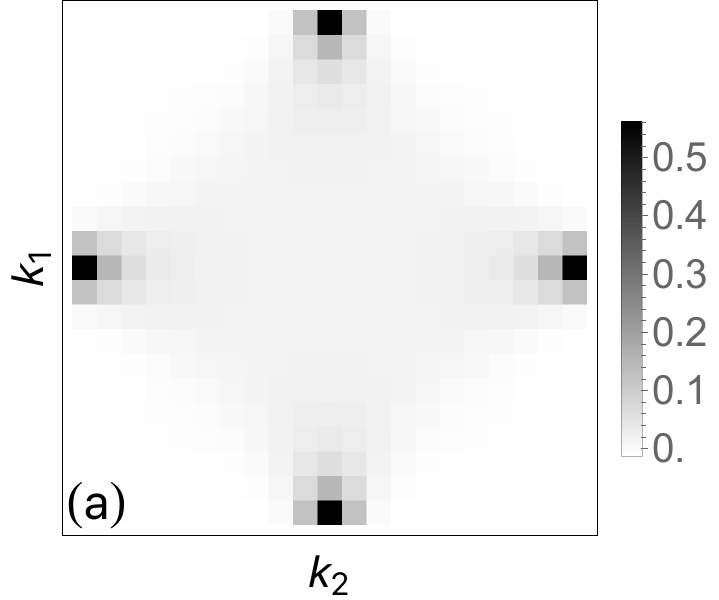}\hfill
\includegraphics[width=0.26\linewidth]{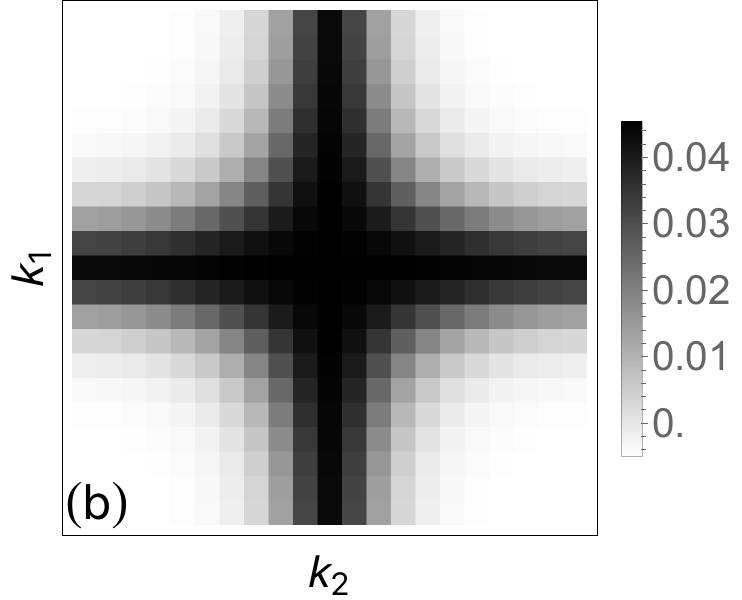}\hfill
\includegraphics[width=0.25\linewidth]{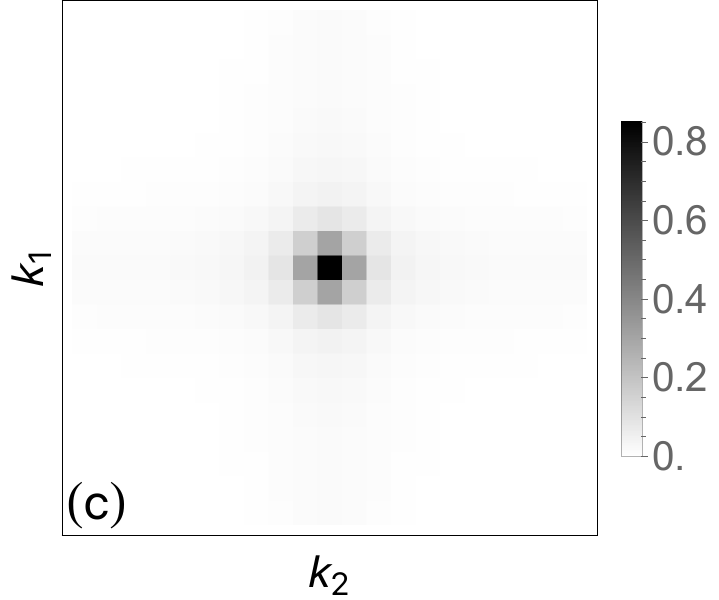}\hfill
\includegraphics[width=0.24\linewidth]{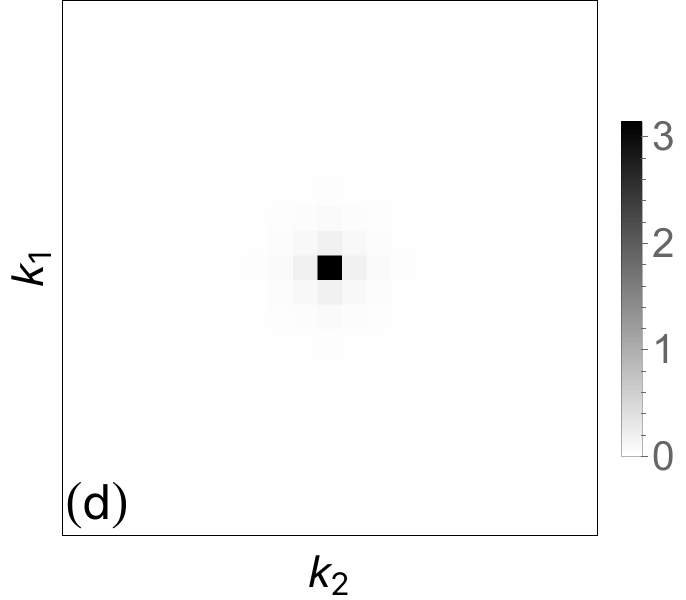}
\caption{Berry-flux density $F_{12}(k)\,d^2k$ on the Brillouin zone $0\le k_\mu\le 2\pi$,
discretized on a $21^2$ mesh,
for the Wilson-Dirac model with $t=b=1$. (a)-(c) show initial distributions for $m=2.2,3.0,3.8$ ($C=+1$). 
(d) shows the common configuration after the trace-type gradient flow.}
\label{f:bc3}
\end{figure*}

\subsection{Invariance of the Chern numbers under gradient flow}

The Chern number is a topological invariant and therefore remains unchanged 
under smooth deformations of the projector. 
Since the gradient flow generates a smooth one-parameter family of projectors, 
we expect the Chern number to be constant along the flow; 
in particular, any fixed point that saturates the Bogomolny bound carries the same Chern number. 
We now show this directly.
Differentiating Eq. (\ref{CheNum}) with respect to $\tau$ yields
\begin{widetext}
\begin{alignat}{1}
\partial_\tau C=\frac{i}{2\pi}\int \varepsilon^{\mu\nu}
\Tr (\partial_\tau P\partial_\mu P\partial_\nu P+P\partial_\mu( \partial_\tau P)\partial_\nu P
+P\partial_\mu P\partial_\nu (\partial_\tau P))d^2k.
\end{alignat}
As discussed in Sec.~\ref{s:actdec}, the gradient flows have the form
$\partial_\tau P= [P,J]$ with an anti-Hermitian $J$ for both actions, whence
$\partial_\mu (\partial_\tau P)=[\partial_\mu P,J]+[P,\partial_\mu J]$. Substituting these, we obtain
\begin{alignat}{1}
\partial_\tau C
&=\frac{i}{2\pi}\int \varepsilon^{\mu\nu}
\Tr \{[P,J]\partial_\mu P\partial_\nu P
+P([\partial_\mu P,J]+[P,\partial_\mu J])\partial_\nu P
+P\partial_\mu P([\partial_\nu P,J]+[P,\partial_\nu J])\}d^2k
\nonumber\\
&=\frac{i}{2\pi}\int \varepsilon^{\mu\nu}
\Tr \left\{([\partial_\mu P\partial_\nu P,P]
+[\partial_\nu P P,\partial_\mu P]+[P\partial_\mu P,\partial_\nu P])J
+(-[\partial_\mu PP,P]
+[P\partial_\mu P,P])\partial_\nu J\right\}d^2k.
\end{alignat}
\end{widetext}
The first commutator vanishes, $[\partial_\mu P\partial_\nu P,P]=0$. Using moreover
$[\partial_\nu PP,\partial_\mu P]=[\partial_\nu P,P\partial_\mu P]$, the second and third commutators cancel, so all terms proportional to $J$ vanish. For the remaining terms containing $\partial_\nu J$, the identities
$[\partial_\mu PP,P]=\partial_\mu PP=(1-P)\partial_\mu P$ and $[P\partial_\mu P,P]=-P\partial_\mu P$ give
\begin{alignat}{1}
\varepsilon^{\mu\nu}&\Tr
(-[\partial_\mu PP,P]
+[P\partial_\mu P,P])\partial_\nu J
\nonumber\\&
=-\partial_\mu (\varepsilon^{\mu\nu}\Tr P\partial_\nu J).
\end{alignat}
Upon integrating over the Brillouin zone  torus, 
which has no boundary, this total divergence vanishes, and therefore $\partial_\tau  C=0$.

\subsection{Trace versus determinant actions}

Since $g_{ab}$ is positive semidefinite, as discussed in Sec. \ref{s:qgt}, 
it has two eigenvalues $\lambda_1,\lambda_2\ge 0$. 
Hence \cite{PhysRevB.90.165139,LEDWITH2021168646,Ozawa:2021vs,PhysRevB.104.045104},
\begin{alignat}{1}
\frac{1}{2}\tr g=\frac{\lambda_1+\lambda_2}{2}\ge \sqrt{\lambda_1\lambda_2}=\det g,
\end{alignat}
with equality iff $\lambda_1=\lambda_2$, equivalently,
\begin{alignat}{1}
g_{11}(k)=g_{22}(k)\quad \text{and}\quad g_{12}(k)=0.
\label{trvsdet}
\end{alignat}
Therefore,  for two actions, we have
\begin{alignat}{1}
S_{\rm t}\ge S_{\rm d},
\label{TrVsDet}
\end{alignat}
and equality holds iff the condition in Eq.~\eqref{trvsdet} is satisfied pointwise for all $k$.

A consequence is worth mentioning. 
Because of Eq.~\eqref{TrVsDet}, reaching the Bogomolny lower bound for the trace functional is 
strictly more demanding than for the determinant  functional: 
to saturate $S_{\rm t}$ according to Eq.~\eqref{TrBog}, 
one must not only saturate the determinant bound \eqref{DetBog}, 
which requires the weighted holomorphic condition \eqref{CauRie}, 
$X_1=i\lambda X_2$ with $X_a=(\partial_a P)P$, but also fulfill the isotropy condition \eqref{trvsdet}. 
Using $g_{ab}=\Re\langle X_a,X_b\rangle$ with $\langle A,B\rangle=\Tr(A^\dagger B)$, this condition gives
\begin{alignat}{1}
g_{12}&=\Re\langle X_1,X_2\rangle
=\frac{i}{2}\,(\lambda-\lambda^*)\,\langle X_2,X_2\rangle=0,\\
g_{11}&=\Re\langle X_1,X_1\rangle
=|\lambda|^2\,\langle X_2,X_2\rangle
=|\lambda|^2 g_{22}.
\end{alignat}
Thus we conclude $\lambda=\pm1$, and the weighted (anti-)holomorphic condition reduces 
to the full (anti-)holomorphic conditions \eqref{Hol}, 
so saturating the trace bound demands both the determinant bound and metric isotropy, consistent with Refs.~\cite{Ozawa:2021vs,PhysRevB.104.045104}.

\section{Gradient flow for the Wilson--Dirac model}\label{s:example}

As a concrete illustration, we apply the gradient flow to the  Wilson-Dirac model,
\begin{alignat}{1}
H(k)=t\,\sigma_\mu\sin k^\mu+\sigma_3\Big(m+b\sum_{\mu=1,2}(\cos k^\mu-1)\Big),
\end{alignat}
where $\sigma_\mu$ stands for the Pauli matrices. 
For $t=b=1$ the occupied-band Chern number is $+1$ for $2<m<4$, $-1$ for $0<m<2$, and $0$ otherwise. 
We initialize the projector $P(k)$ from the occupied eigenstate of $H(k)$ and evolve it under the trace flow \eqref{TraGraFlo}.

Figures\,\ref{f:bc3}(a)--\ref{f:bc3}(c) display the Berry flux density $F_{12}(k)\,d^2k$ for three representative masses 
in the same topological sector $C=+1$, 
and Fig. \ref{f:bc3}(d) shows the common outcome after the trace-flow evolution from those initial conditions. 
Here, the Berry curvature is evaluated by the gauge-invariant discrete plaquette formulation in Ref. \cite{FHS05}.
A salient feature is the large dynamic-range change along the flow: 
initially the curvature is weak and spread out; its weight extends along the reciprocal axes and 
is biased away from $(\pi,0)$ and $(0,\pi)$ toward the Brillouin zone center, $(\pi,\pi)$, whereas after the flow it is strongly concentrated into a single, nearly rotationally symmetric lump at $(\pi,\pi)$. 
Since the total flux is topological and conserved, the sharpening appears as a strong increase of 
the peak value.
This makes the numerical evolution stiff; we therefore employ smaller time steps 
as the bound is approached.

Overall, these examples show that, within a fixed topological phase, 
different parameter choices of the same lattice Hamiltonian flow to a canonical, 
low-complexity representative: the Berry curvature is sharpened and concentrated into a single, nearly isotropic peak,
while the Chern number remains unchanged. 
This supports our broader claim that the gradient flow provides a practical route to simplify models 
within a given topological class without altering their topological charge.

\section{Summary and discussion}\label{s:summary}

We developed a variational framework for projector fields $P(k)$ 
that relies solely on quantum-geometric data of an isolated set of Bloch bands. 
We analyzed two natural functionals in detail: a trace-type action and a determinant-type “quantum volume’’ action constructed from the quantum metric. 
For both, we derived projector-preserving gradient flows in closed form, proved their monotonic decrease, 
and established Bogomolny-type lower bounds tied to the Chern number. 
The bounds are saturated by (anti-)holomorphic projector configurations, 
and the Chern number is strictly conserved along the flows. 
We also showed that the trace functional dominates the determinant, 
with equality precisely when the metric is isotropic in the orthonormal frame. 
A concrete case study with the Wilson-Dirac model illustrated how distinct initial curvature profiles 
within the same topological sector are driven by the flow toward canonical, 
low-complexity representatives without changing the Chern number.

Conceptually, the gradient flows act as geometry-only smoothing dynamics on the space of Bloch states, 
projecting arbitrary deformations onto the tangent of the projector manifold 
and removing model-specific anisotropies while retaining topological charge. 
The Bogomolny equations take a simple (anti-)holomorphic form in an orthonormal frame, 
clarifying the role of complex structure behind the saturating configurations. 
For the determinant functional, the weighted holomorphic condition provides 
a natural interpolation to the same (anti-)holomorphic fixed points.

There are several  open directions. 
First,  the uniqueness of the limiting canonical representative 
remains to be characterized; constraints from lattice symmetries or additional quantum numbers 
could be incorporated to refine the flow. 
Second, while our numerics keep $P^2=P$ exactly by construction, 
discretization effects on the Brillouin zone, and their impact on action-decay rates and curvature-lump shapes, 
warrant a systematic convergence study. 
Finally, it is natural to ask how the present framework adapts to disordered settings, 
to interacting systems where the ground-state projector is replaced by a many-body density projector, 
or to non-Hermitian band structures.

\acknowledgements
The author thanks H. Suzuki and O. Morikawa for valuable discussions on gradient flows.
This work was supported in part by a Grant-in-Aid for Scientific Research (Grant No.~22K03448) 
from the Japan Society for the Promotion of Science.


\begin{thebibliography}{29}%
\makeatletter
\providecommand \@ifxundefined [1]{%
 \@ifx{#1\undefined}
}%
\providecommand \@ifnum [1]{%
 \ifnum #1\expandafter \@firstoftwo
 \else \expandafter \@secondoftwo
 \fi
}%
\providecommand \@ifx [1]{%
 \ifx #1\expandafter \@firstoftwo
 \else \expandafter \@secondoftwo
 \fi
}%
\providecommand \natexlab [1]{#1}%
\providecommand \enquote  [1]{``#1''}%
\providecommand \bibnamefont  [1]{#1}%
\providecommand \bibfnamefont [1]{#1}%
\providecommand \citenamefont [1]{#1}%
\providecommand \href@noop [0]{\@secondoftwo}%
\providecommand \href [0]{\begingroup \@sanitize@url \@href}%
\providecommand \@href[1]{\@@startlink{#1}\@@href}%
\providecommand \@@href[1]{\endgroup#1\@@endlink}%
\providecommand \@sanitize@url [0]{\catcode `\\12\catcode `\$12\catcode
  `\&12\catcode `\#12\catcode `\^12\catcode `\_12\catcode `\%12\relax}%
\providecommand \@@startlink[1]{}%
\providecommand \@@endlink[0]{}%
\providecommand \url  [0]{\begingroup\@sanitize@url \@url }%
\providecommand \@url [1]{\endgroup\@href {#1}{\urlprefix }}%
\providecommand \urlprefix  [0]{URL }%
\providecommand \Eprint [0]{\href }%
\providecommand \doibase [0]{https://doi.org/}%
\providecommand \selectlanguage [0]{\@gobble}%
\providecommand \bibinfo  [0]{\@secondoftwo}%
\providecommand \bibfield  [0]{\@secondoftwo}%
\providecommand \translation [1]{[#1]}%
\providecommand \BibitemOpen [0]{}%
\providecommand \bibitemStop [0]{}%
\providecommand \bibitemNoStop [0]{.\EOS\space}%
\providecommand \EOS [0]{\spacefactor3000\relax}%
\providecommand \BibitemShut  [1]{\csname bibitem#1\endcsname}%
\let\auto@bib@innerbib\@empty
\bibitem [{\citenamefont {Provost}\ and\ \citenamefont
  {Vallee}(1980)}]{Provost:1980aa}%
  \BibitemOpen
  \bibfield  {author} {\bibinfo {author} {\bibfnamefont {J.~P.}\ \bibnamefont
  {Provost}}\ and\ \bibinfo {author} {\bibfnamefont {G.}~\bibnamefont
  {Vallee}},\ }\bibfield  {title} {\bibinfo {title} {Riemannian structure on
  manifolds of quantum states},\ }\href {https://doi.org/10.1007/BF02193559}
  {\bibfield  {journal} {\bibinfo  {journal} {Communications in Mathematical
  Physics}\ }\textbf {\bibinfo {volume} {76}},\ \bibinfo {pages} {289}
  (\bibinfo {year} {1980})}\BibitemShut {NoStop}%
\bibitem [{\citenamefont {Thouless}\ \emph {et~al.}(1982)\citenamefont
  {Thouless}, \citenamefont {Kohmoto}, \citenamefont {Nightingale},\ and\
  \citenamefont {den Nijs}}]{Thouless:1982uq}%
  \BibitemOpen
  \bibfield  {author} {\bibinfo {author} {\bibfnamefont {D.~J.}\ \bibnamefont
  {Thouless}}, \bibinfo {author} {\bibfnamefont {M.}~\bibnamefont {Kohmoto}},
  \bibinfo {author} {\bibfnamefont {M.~P.}\ \bibnamefont {Nightingale}},\ and\
  \bibinfo {author} {\bibfnamefont {M.}~\bibnamefont {den Nijs}},\ }\bibfield
  {title} {\bibinfo {title} {Quantized hall conductance in a two-dimensional
  periodic potential},\ }\href
  {http://link.aps.org/doi/10.1103/PhysRevLett.49.405} {\bibfield  {journal}
  {\bibinfo  {journal} {Physical Review Letters}\ }\textbf {\bibinfo {volume}
  {49}},\ \bibinfo {pages} {405} (\bibinfo {year} {1982})}\BibitemShut
  {NoStop}%
\bibitem [{\citenamefont {Kohmoto}(1985)}]{kohmoto:85}%
  \BibitemOpen
  \bibfield  {author} {\bibinfo {author} {\bibfnamefont {M.}~\bibnamefont
  {Kohmoto}},\ }\bibfield  {title} {\bibinfo {title} {Topological invariant and
  the quantization of the hall conductance},\ }\href@noop {} {\bibfield
  {journal} {\bibinfo  {journal} {Annals of Physics}\ }\textbf {\bibinfo
  {volume} {160}},\ \bibinfo {pages} {343} (\bibinfo {year}
  {1985})}\BibitemShut {NoStop}%
\bibitem [{\citenamefont {Kolodrubetz}\ \emph {et~al.}(2017)\citenamefont
  {Kolodrubetz}, \citenamefont {Sels}, \citenamefont {Mehta},\ and\
  \citenamefont {Polkovnikov}}]{Kolodrubetz:2017aa}%
  \BibitemOpen
  \bibfield  {author} {\bibinfo {author} {\bibfnamefont {M.}~\bibnamefont
  {Kolodrubetz}}, \bibinfo {author} {\bibfnamefont {D.}~\bibnamefont {Sels}},
  \bibinfo {author} {\bibfnamefont {P.}~\bibnamefont {Mehta}},\ and\ \bibinfo
  {author} {\bibfnamefont {A.}~\bibnamefont {Polkovnikov}},\ }\bibfield
  {title} {\bibinfo {title} {Geometry and non-adiabatic response in quantum and
  classical systems},\ }\bibfield  {booktitle} {\emph {\bibinfo {booktitle}
  {Geometry and non-adiabatic response in quantum and classical systems}},\
  }\href {https://doi.org/https://doi.org/10.1016/j.physrep.2017.07.001}
  {\bibfield  {journal} {\bibinfo  {journal} {Physics Reports}\ }\textbf
  {\bibinfo {volume} {697}},\ \bibinfo {pages} {1} (\bibinfo {year}
  {2017})}\BibitemShut {NoStop}%
\bibitem [{\citenamefont {Peotta}\ and\ \citenamefont
  {T{\"o}rm{\"a}}(2015)}]{Peotta:2015aa}%
  \BibitemOpen
  \bibfield  {author} {\bibinfo {author} {\bibfnamefont {S.}~\bibnamefont
  {Peotta}}\ and\ \bibinfo {author} {\bibfnamefont {P.}~\bibnamefont
  {T{\"o}rm{\"a}}},\ }\bibfield  {title} {\bibinfo {title} {Superfluidity in
  topologically nontrivial flat bands},\ }\href
  {https://doi.org/10.1038/ncomms9944} {\bibfield  {journal} {\bibinfo
  {journal} {Nature Communications}\ }\textbf {\bibinfo {volume} {6}},\
  \bibinfo {pages} {8944 EP } (\bibinfo {year} {2015})}\BibitemShut {NoStop}%
\bibitem [{\citenamefont {Liang}\ \emph {et~al.}(2017)\citenamefont {Liang},
  \citenamefont {Vanhala}, \citenamefont {Peotta}, \citenamefont {Siro},
  \citenamefont {Harju},\ and\ \citenamefont {T\"orm\"a}}]{PhysRevB.95.024515}%
  \BibitemOpen
  \bibfield  {author} {\bibinfo {author} {\bibfnamefont {L.}~\bibnamefont
  {Liang}}, \bibinfo {author} {\bibfnamefont {T.~I.}\ \bibnamefont {Vanhala}},
  \bibinfo {author} {\bibfnamefont {S.}~\bibnamefont {Peotta}}, \bibinfo
  {author} {\bibfnamefont {T.}~\bibnamefont {Siro}}, \bibinfo {author}
  {\bibfnamefont {A.}~\bibnamefont {Harju}},\ and\ \bibinfo {author}
  {\bibfnamefont {P.}~\bibnamefont {T\"orm\"a}},\ }\bibfield  {title} {\bibinfo
  {title} {Band geometry, berry curvature, and superfluid weight},\ }\href
  {https://doi.org/10.1103/PhysRevB.95.024515} {\bibfield  {journal} {\bibinfo
  {journal} {Phys. Rev. B}\ }\textbf {\bibinfo {volume} {95}},\ \bibinfo
  {pages} {024515} (\bibinfo {year} {2017})}\BibitemShut {NoStop}%
\bibitem [{\citenamefont {Marzari}\ and\ \citenamefont
  {Vanderbilt}(1997)}]{Marzari:1997aa}%
  \BibitemOpen
  \bibfield  {author} {\bibinfo {author} {\bibfnamefont {N.}~\bibnamefont
  {Marzari}}\ and\ \bibinfo {author} {\bibfnamefont {D.}~\bibnamefont
  {Vanderbilt}},\ }\bibfield  {title} {\bibinfo {title} {Maximally localized
  generalized wannier functions for composite energy bands},\ }\href
  {http://link.aps.org/doi/10.1103/PhysRevB.56.12847} {\bibfield  {journal}
  {\bibinfo  {journal} {Physical Review B}\ }\textbf {\bibinfo {volume} {56}},\
  \bibinfo {pages} {12847} (\bibinfo {year} {1997})}\BibitemShut {NoStop}%
\bibitem [{\citenamefont {Resta}\ and\ \citenamefont
  {Sorella}(1999)}]{PhysRevLett.82.370}%
  \BibitemOpen
  \bibfield  {author} {\bibinfo {author} {\bibfnamefont {R.}~\bibnamefont
  {Resta}}\ and\ \bibinfo {author} {\bibfnamefont {S.}~\bibnamefont
  {Sorella}},\ }\bibfield  {title} {\bibinfo {title} {Electron localization in
  the insulating state},\ }\href {https://doi.org/10.1103/PhysRevLett.82.370}
  {\bibfield  {journal} {\bibinfo  {journal} {Phys. Rev. Lett.}\ }\textbf
  {\bibinfo {volume} {82}},\ \bibinfo {pages} {370} (\bibinfo {year}
  {1999})}\BibitemShut {NoStop}%
\bibitem [{\citenamefont {Sipe}\ and\ \citenamefont
  {Shkrebtii}(2000)}]{PhysRevB.61.5337}%
  \BibitemOpen
  \bibfield  {author} {\bibinfo {author} {\bibfnamefont {J.~E.}\ \bibnamefont
  {Sipe}}\ and\ \bibinfo {author} {\bibfnamefont {A.~I.}\ \bibnamefont
  {Shkrebtii}},\ }\bibfield  {title} {\bibinfo {title} {Second-order optical
  response in semiconductors},\ }\href
  {https://doi.org/10.1103/PhysRevB.61.5337} {\bibfield  {journal} {\bibinfo
  {journal} {Phys. Rev. B}\ }\textbf {\bibinfo {volume} {61}},\ \bibinfo
  {pages} {5337} (\bibinfo {year} {2000})}\BibitemShut {NoStop}%
\bibitem [{\citenamefont {Morimoto}\ and\ \citenamefont
  {Nagaosa}(2016)}]{doi:10.1126/sciadv.1501524}%
  \BibitemOpen
  \bibfield  {author} {\bibinfo {author} {\bibfnamefont {T.}~\bibnamefont
  {Morimoto}}\ and\ \bibinfo {author} {\bibfnamefont {N.}~\bibnamefont
  {Nagaosa}},\ }\bibfield  {title} {\bibinfo {title} {Topological nature of
  nonlinear optical effects in solids},\ }\href
  {https://doi.org/10.1126/sciadv.1501524} {\bibfield  {journal} {\bibinfo
  {journal} {Science Advances}\ }\textbf {\bibinfo {volume} {2}},\ \bibinfo
  {pages} {e1501524} (\bibinfo {year} {2016})},\ \Eprint
  {https://arxiv.org/abs/https://www.science.org/doi/pdf/10.1126/sciadv.1501524}
  {https://www.science.org/doi/pdf/10.1126/sciadv.1501524} \BibitemShut
  {NoStop}%
\bibitem [{\citenamefont {Ahn}\ \emph {et~al.}(2020)\citenamefont {Ahn},
  \citenamefont {Guo},\ and\ \citenamefont {Nagaosa}}]{PhysRevX.10.041041}%
  \BibitemOpen
  \bibfield  {author} {\bibinfo {author} {\bibfnamefont {J.}~\bibnamefont
  {Ahn}}, \bibinfo {author} {\bibfnamefont {G.-Y.}\ \bibnamefont {Guo}},\ and\
  \bibinfo {author} {\bibfnamefont {N.}~\bibnamefont {Nagaosa}},\ }\bibfield
  {title} {\bibinfo {title} {Low-frequency divergence and quantum geometry of
  the bulk photovoltaic effect in topological semimetals},\ }\href
  {https://doi.org/10.1103/PhysRevX.10.041041} {\bibfield  {journal} {\bibinfo
  {journal} {Phys. Rev. X}\ }\textbf {\bibinfo {volume} {10}},\ \bibinfo
  {pages} {041041} (\bibinfo {year} {2020})}\BibitemShut {NoStop}%
\bibitem [{\citenamefont {Bhalla}\ \emph {et~al.}(2022)\citenamefont {Bhalla},
  \citenamefont {Das}, \citenamefont {Culcer},\ and\ \citenamefont
  {Agarwal}}]{PhysRevLett.129.227401}%
  \BibitemOpen
  \bibfield  {author} {\bibinfo {author} {\bibfnamefont {P.}~\bibnamefont
  {Bhalla}}, \bibinfo {author} {\bibfnamefont {K.}~\bibnamefont {Das}},
  \bibinfo {author} {\bibfnamefont {D.}~\bibnamefont {Culcer}},\ and\ \bibinfo
  {author} {\bibfnamefont {A.}~\bibnamefont {Agarwal}},\ }\bibfield  {title}
  {\bibinfo {title} {Resonant second-harmonic generation as a probe of quantum
  geometry},\ }\href {https://doi.org/10.1103/PhysRevLett.129.227401}
  {\bibfield  {journal} {\bibinfo  {journal} {Phys. Rev. Lett.}\ }\textbf
  {\bibinfo {volume} {129}},\ \bibinfo {pages} {227401} (\bibinfo {year}
  {2022})}\BibitemShut {NoStop}%
\bibitem [{\citenamefont {Sodemann}\ and\ \citenamefont
  {Fu}(2015)}]{PhysRevLett.115.216806}%
  \BibitemOpen
  \bibfield  {author} {\bibinfo {author} {\bibfnamefont {I.}~\bibnamefont
  {Sodemann}}\ and\ \bibinfo {author} {\bibfnamefont {L.}~\bibnamefont {Fu}},\
  }\bibfield  {title} {\bibinfo {title} {Quantum nonlinear hall effect induced
  by berry curvature dipole in time-reversal invariant materials},\ }\href
  {https://doi.org/10.1103/PhysRevLett.115.216806} {\bibfield  {journal}
  {\bibinfo  {journal} {Phys. Rev. Lett.}\ }\textbf {\bibinfo {volume} {115}},\
  \bibinfo {pages} {216806} (\bibinfo {year} {2015})}\BibitemShut {NoStop}%
\bibitem [{\citenamefont {Das}\ \emph {et~al.}(2023)\citenamefont {Das},
  \citenamefont {Lahiri}, \citenamefont {Atencia}, \citenamefont {Culcer},\
  and\ \citenamefont {Agarwal}}]{PhysRevB.108.L201405}%
  \BibitemOpen
  \bibfield  {author} {\bibinfo {author} {\bibfnamefont {K.}~\bibnamefont
  {Das}}, \bibinfo {author} {\bibfnamefont {S.}~\bibnamefont {Lahiri}},
  \bibinfo {author} {\bibfnamefont {R.~B.}\ \bibnamefont {Atencia}}, \bibinfo
  {author} {\bibfnamefont {D.}~\bibnamefont {Culcer}},\ and\ \bibinfo {author}
  {\bibfnamefont {A.}~\bibnamefont {Agarwal}},\ }\bibfield  {title} {\bibinfo
  {title} {Intrinsic nonlinear conductivities induced by the quantum metric},\
  }\href {https://doi.org/10.1103/PhysRevB.108.L201405} {\bibfield  {journal}
  {\bibinfo  {journal} {Phys. Rev. B}\ }\textbf {\bibinfo {volume} {108}},\
  \bibinfo {pages} {L201405} (\bibinfo {year} {2023})}\BibitemShut {NoStop}%
\bibitem [{\citenamefont {Gao}\ \emph {et~al.}(2023)\citenamefont {Gao},
  \citenamefont {Liu}, \citenamefont {Qiu}, \citenamefont {Ghosh},
  \citenamefont {Trevisan}, \citenamefont {Onishi}, \citenamefont {Hu},
  \citenamefont {Qian}, \citenamefont {Tien}, \citenamefont {Chen},
  \citenamefont {Huang}, \citenamefont {B{\'e}rub{\'e}}, \citenamefont {Li},
  \citenamefont {Tzschaschel}, \citenamefont {Dinh}, \citenamefont {Sun},
  \citenamefont {Ho}, \citenamefont {Lien}, \citenamefont {Singh},
  \citenamefont {Watanabe}, \citenamefont {Taniguchi}, \citenamefont {Bell},
  \citenamefont {Lin}, \citenamefont {Chang}, \citenamefont {Du}, \citenamefont
  {Bansil}, \citenamefont {Fu}, \citenamefont {Ni}, \citenamefont {Orth},
  \citenamefont {Ma},\ and\ \citenamefont {Xu}}]{doi:10.1126/science.adf1506}%
  \BibitemOpen
  \bibfield  {author} {\bibinfo {author} {\bibfnamefont {A.}~\bibnamefont
  {Gao}}, \bibinfo {author} {\bibfnamefont {Y.-F.}\ \bibnamefont {Liu}},
  \bibinfo {author} {\bibfnamefont {J.-X.}\ \bibnamefont {Qiu}}, \bibinfo
  {author} {\bibfnamefont {B.}~\bibnamefont {Ghosh}}, \bibinfo {author}
  {\bibfnamefont {T.~V.}\ \bibnamefont {Trevisan}}, \bibinfo {author}
  {\bibfnamefont {Y.}~\bibnamefont {Onishi}}, \bibinfo {author} {\bibfnamefont
  {C.}~\bibnamefont {Hu}}, \bibinfo {author} {\bibfnamefont {T.}~\bibnamefont
  {Qian}}, \bibinfo {author} {\bibfnamefont {H.-J.}\ \bibnamefont {Tien}},
  \bibinfo {author} {\bibfnamefont {S.-W.}\ \bibnamefont {Chen}}, \bibinfo
  {author} {\bibfnamefont {M.}~\bibnamefont {Huang}}, \bibinfo {author}
  {\bibfnamefont {D.}~\bibnamefont {B{\'e}rub{\'e}}}, \bibinfo {author}
  {\bibfnamefont {H.}~\bibnamefont {Li}}, \bibinfo {author} {\bibfnamefont
  {C.}~\bibnamefont {Tzschaschel}}, \bibinfo {author} {\bibfnamefont
  {T.}~\bibnamefont {Dinh}}, \bibinfo {author} {\bibfnamefont {Z.}~\bibnamefont
  {Sun}}, \bibinfo {author} {\bibfnamefont {S.-C.}\ \bibnamefont {Ho}},
  \bibinfo {author} {\bibfnamefont {S.-W.}\ \bibnamefont {Lien}}, \bibinfo
  {author} {\bibfnamefont {B.}~\bibnamefont {Singh}}, \bibinfo {author}
  {\bibfnamefont {K.}~\bibnamefont {Watanabe}}, \bibinfo {author}
  {\bibfnamefont {T.}~\bibnamefont {Taniguchi}}, \bibinfo {author}
  {\bibfnamefont {D.~C.}\ \bibnamefont {Bell}}, \bibinfo {author}
  {\bibfnamefont {H.}~\bibnamefont {Lin}}, \bibinfo {author} {\bibfnamefont
  {T.-R.}\ \bibnamefont {Chang}}, \bibinfo {author} {\bibfnamefont {C.~R.}\
  \bibnamefont {Du}}, \bibinfo {author} {\bibfnamefont {A.}~\bibnamefont
  {Bansil}}, \bibinfo {author} {\bibfnamefont {L.}~\bibnamefont {Fu}}, \bibinfo
  {author} {\bibfnamefont {N.}~\bibnamefont {Ni}}, \bibinfo {author}
  {\bibfnamefont {P.~P.}\ \bibnamefont {Orth}}, \bibinfo {author}
  {\bibfnamefont {Q.}~\bibnamefont {Ma}},\ and\ \bibinfo {author}
  {\bibfnamefont {S.-Y.}\ \bibnamefont {Xu}},\ }\bibfield  {title} {\bibinfo
  {title} {Quantum metric nonlinear hall effect in a topological
  antiferromagnetic heterostructure},\ }\href
  {https://doi.org/10.1126/science.adf1506} {\bibfield  {journal} {\bibinfo
  {journal} {Science}\ }\textbf {\bibinfo {volume} {381}},\ \bibinfo {pages}
  {181} (\bibinfo {year} {2023})},\ \Eprint
  {https://arxiv.org/abs/https://www.science.org/doi/pdf/10.1126/science.adf1506}
  {https://www.science.org/doi/pdf/10.1126/science.adf1506} \BibitemShut
  {NoStop}%
\bibitem [{\citenamefont {Souza}\ \emph {et~al.}(2000)\citenamefont {Souza},
  \citenamefont {Wilkens},\ and\ \citenamefont {Martin}}]{PhysRevB.62.1666}%
  \BibitemOpen
  \bibfield  {author} {\bibinfo {author} {\bibfnamefont {I.}~\bibnamefont
  {Souza}}, \bibinfo {author} {\bibfnamefont {T.}~\bibnamefont {Wilkens}},\
  and\ \bibinfo {author} {\bibfnamefont {R.~M.}\ \bibnamefont {Martin}},\
  }\bibfield  {title} {\bibinfo {title} {Polarization and localization in
  insulators: Generating function approach},\ }\href
  {https://doi.org/10.1103/PhysRevB.62.1666} {\bibfield  {journal} {\bibinfo
  {journal} {Phys. Rev. B}\ }\textbf {\bibinfo {volume} {62}},\ \bibinfo
  {pages} {1666} (\bibinfo {year} {2000})}\BibitemShut {NoStop}%
\bibitem [{\citenamefont {Onishi}\ and\ \citenamefont
  {Fu}(2025)}]{PhysRevResearch.7.023158}%
  \BibitemOpen
  \bibfield  {author} {\bibinfo {author} {\bibfnamefont {Y.}~\bibnamefont
  {Onishi}}\ and\ \bibinfo {author} {\bibfnamefont {L.}~\bibnamefont {Fu}},\
  }\bibfield  {title} {\bibinfo {title} {Quantum weight: A fundamental property
  of quantum many-body systems},\ }\href
  {https://doi.org/10.1103/PhysRevResearch.7.023158} {\bibfield  {journal}
  {\bibinfo  {journal} {Phys. Rev. Res.}\ }\textbf {\bibinfo {volume} {7}},\
  \bibinfo {pages} {023158} (\bibinfo {year} {2025})}\BibitemShut {NoStop}%
\bibitem [{\citenamefont {Het\'enyi}\ and\ \citenamefont
  {L\'evay}(2023)}]{PhysRevA.108.032218}%
  \BibitemOpen
  \bibfield  {author} {\bibinfo {author} {\bibfnamefont {B.}~\bibnamefont
  {Het\'enyi}}\ and\ \bibinfo {author} {\bibfnamefont {P.}~\bibnamefont
  {L\'evay}},\ }\bibfield  {title} {\bibinfo {title} {Fluctuations, uncertainty
  relations, and the geometry of quantum state manifolds},\ }\href
  {https://doi.org/10.1103/PhysRevA.108.032218} {\bibfield  {journal} {\bibinfo
   {journal} {Phys. Rev. A}\ }\textbf {\bibinfo {volume} {108}},\ \bibinfo
  {pages} {032218} (\bibinfo {year} {2023})}\BibitemShut {NoStop}%
\bibitem [{\citenamefont {Ledwith}\ \emph {et~al.}(2021)\citenamefont
  {Ledwith}, \citenamefont {Khalaf},\ and\ \citenamefont
  {Vishwanath}}]{LEDWITH2021168646}%
  \BibitemOpen
  \bibfield  {author} {\bibinfo {author} {\bibfnamefont {P.~J.}\ \bibnamefont
  {Ledwith}}, \bibinfo {author} {\bibfnamefont {E.}~\bibnamefont {Khalaf}},\
  and\ \bibinfo {author} {\bibfnamefont {A.}~\bibnamefont {Vishwanath}},\
  }\bibfield  {title} {\bibinfo {title} {Strong coupling theory of magic-angle
  graphene: A pedagogical introduction},\ }\href
  {https://doi.org/https://doi.org/10.1016/j.aop.2021.168646} {\bibfield
  {journal} {\bibinfo  {journal} {Annals of Physics}\ }\textbf {\bibinfo
  {volume} {435}},\ \bibinfo {pages} {168646} (\bibinfo {year} {2021})},\
  \bibinfo {note} {special issue on Philip W. Anderson}\BibitemShut {NoStop}%
\bibitem [{\citenamefont {Ozawa}\ and\ \citenamefont
  {Mera}(2021)}]{Ozawa:2021vs}%
  \BibitemOpen
  \bibfield  {author} {\bibinfo {author} {\bibfnamefont {T.}~\bibnamefont
  {Ozawa}}\ and\ \bibinfo {author} {\bibfnamefont {B.}~\bibnamefont {Mera}},\
  }\bibfield  {title} {\bibinfo {title} {Relations between topology and the
  quantum metric for chern insulators},\ }\href
  {https://doi.org/10.1103/PhysRevB.104.045103} {\bibfield  {journal} {\bibinfo
   {journal} {Physical Review B}\ }\textbf {\bibinfo {volume} {104}},\ \bibinfo
  {pages} {045103} (\bibinfo {year} {2021})}\BibitemShut {NoStop}%
\bibitem [{\citenamefont {Mera}\ and\ \citenamefont
  {Ozawa}(2021)}]{PhysRevB.104.045104}%
  \BibitemOpen
  \bibfield  {author} {\bibinfo {author} {\bibfnamefont {B.}~\bibnamefont
  {Mera}}\ and\ \bibinfo {author} {\bibfnamefont {T.}~\bibnamefont {Ozawa}},\
  }\bibfield  {title} {\bibinfo {title} {K\"ahler geometry and chern
  insulators: Relations between topology and the quantum metric},\ }\href
  {https://doi.org/10.1103/PhysRevB.104.045104} {\bibfield  {journal} {\bibinfo
   {journal} {Phys. Rev. B}\ }\textbf {\bibinfo {volume} {104}},\ \bibinfo
  {pages} {045104} (\bibinfo {year} {2021})}\BibitemShut {NoStop}%
\bibitem [{\citenamefont {Liu}\ \emph {et~al.}(2025)\citenamefont {Liu},
  \citenamefont {Mera}, \citenamefont {Fujimoto}, \citenamefont {Ozawa},\ and\
  \citenamefont {Wang}}]{1zg9-qbd6}%
  \BibitemOpen
  \bibfield  {author} {\bibinfo {author} {\bibfnamefont {Z.}~\bibnamefont
  {Liu}}, \bibinfo {author} {\bibfnamefont {B.}~\bibnamefont {Mera}}, \bibinfo
  {author} {\bibfnamefont {M.}~\bibnamefont {Fujimoto}}, \bibinfo {author}
  {\bibfnamefont {T.}~\bibnamefont {Ozawa}},\ and\ \bibinfo {author}
  {\bibfnamefont {J.}~\bibnamefont {Wang}},\ }\bibfield  {title} {\bibinfo
  {title} {Theory of generalized landau levels and its implications for
  non-abelian states},\ }\href {https://doi.org/10.1103/1zg9-qbd6} {\bibfield
  {journal} {\bibinfo  {journal} {Phys. Rev. X}\ }\textbf {\bibinfo {volume}
  {15}},\ \bibinfo {pages} {031019} (\bibinfo {year} {2025})}\BibitemShut
  {NoStop}%
\bibitem [{\citenamefont {Onishi}\ \emph {et~al.}(2025)\citenamefont {Onishi},
  \citenamefont {Avdoshkin},\ and\ \citenamefont {Fu}}]{8ng1-bwf6}%
  \BibitemOpen
  \bibfield  {author} {\bibinfo {author} {\bibfnamefont {Y.}~\bibnamefont
  {Onishi}}, \bibinfo {author} {\bibfnamefont {A.}~\bibnamefont {Avdoshkin}},\
  and\ \bibinfo {author} {\bibfnamefont {L.}~\bibnamefont {Fu}},\ }\bibfield
  {title} {\bibinfo {title} {Geometric bound on the structure factor and a
  harmonic condition on band geometry},\ }\href
  {https://doi.org/10.1103/8ng1-bwf6} {\bibfield  {journal} {\bibinfo
  {journal} {Phys. Rev. B}\ }\textbf {\bibinfo {volume} {112}},\ \bibinfo
  {pages} {035158} (\bibinfo {year} {2025})}\BibitemShut {NoStop}%
\bibitem [{\citenamefont {Mitscherling}\ \emph {et~al.}(2025)\citenamefont
  {Mitscherling}, \citenamefont {Avdoshkin},\ and\ \citenamefont
  {Moore}}]{qscv-qxqt}%
  \BibitemOpen
  \bibfield  {author} {\bibinfo {author} {\bibfnamefont {J.}~\bibnamefont
  {Mitscherling}}, \bibinfo {author} {\bibfnamefont {A.}~\bibnamefont
  {Avdoshkin}},\ and\ \bibinfo {author} {\bibfnamefont {J.~E.}\ \bibnamefont
  {Moore}},\ }\bibfield  {title} {\bibinfo {title} {Gauge-invariant projector
  calculus for quantum state geometry and applications to observables in
  crystals},\ }\href {https://doi.org/10.1103/qscv-qxqt} {\bibfield  {journal}
  {\bibinfo  {journal} {Phys. Rev. B}\ }\textbf {\bibinfo {volume} {112}},\
  \bibinfo {pages} {085104} (\bibinfo {year} {2025})}\BibitemShut {NoStop}%
\bibitem [{\citenamefont {Shinada}\ and\ \citenamefont
  {Nagaosa}(2025)}]{qxbl-qd4f}%
  \BibitemOpen
  \bibfield  {author} {\bibinfo {author} {\bibfnamefont {K.}~\bibnamefont
  {Shinada}}\ and\ \bibinfo {author} {\bibfnamefont {N.}~\bibnamefont
  {Nagaosa}},\ }\bibfield  {title} {\bibinfo {title} {Quantum geometric bounds
  for observables: Linear responses, drude weight, and orbital magnetization},\
  }\href {https://doi.org/10.1103/qxbl-qd4f} {\bibfield  {journal} {\bibinfo
  {journal} {Phys. Rev. B}\ }\textbf {\bibinfo {volume} {112}},\ \bibinfo
  {pages} {155158} (\bibinfo {year} {2025})}\BibitemShut {NoStop}%
\bibitem [{\citenamefont {Yu}\ \emph {et~al.}(2025)\citenamefont {Yu},
  \citenamefont {Lian},\ and\ \citenamefont
  {Ryu}}]{yu2025wilsonloopidealbandsgeneralidealization}%
  \BibitemOpen
  \bibfield  {author} {\bibinfo {author} {\bibfnamefont {J.}~\bibnamefont
  {Yu}}, \bibinfo {author} {\bibfnamefont {B.}~\bibnamefont {Lian}},\ and\
  \bibinfo {author} {\bibfnamefont {S.}~\bibnamefont {Ryu}},\ }\href
  {https://arxiv.org/abs/2509.05410} {\bibinfo {title} {Wilson-loop-ideal bands
  and general idealization}} (\bibinfo {year} {2025}),\ \Eprint
  {https://arxiv.org/abs/2509.05410} {arXiv:2509.05410 [cond-mat.mes-hall]}
  \BibitemShut {NoStop}%
\bibitem [{\citenamefont {Manton}\ and\ \citenamefont
  {Sutcliffe}(2004)}]{MantonSutcliffe2004}%
  \BibitemOpen
  \bibfield  {author} {\bibinfo {author} {\bibfnamefont {N.}~\bibnamefont
  {Manton}}\ and\ \bibinfo {author} {\bibfnamefont {P.}~\bibnamefont
  {Sutcliffe}},\ }\href@noop {} {\emph {\bibinfo {title} {Topological
  Solitons}}},\ Cambridge Monographs on Mathematical Physics\ (\bibinfo
  {publisher} {Cambridge University Press},\ \bibinfo {address} {Cambridge},\
  \bibinfo {year} {2004})\BibitemShut {NoStop}%
\bibitem [{\citenamefont {Roy}(2014)}]{PhysRevB.90.165139}%
  \BibitemOpen
  \bibfield  {author} {\bibinfo {author} {\bibfnamefont {R.}~\bibnamefont
  {Roy}},\ }\bibfield  {title} {\bibinfo {title} {Band geometry of fractional
  topological insulators},\ }\href {https://doi.org/10.1103/PhysRevB.90.165139}
  {\bibfield  {journal} {\bibinfo  {journal} {Phys. Rev. B}\ }\textbf {\bibinfo
  {volume} {90}},\ \bibinfo {pages} {165139} (\bibinfo {year}
  {2014})}\BibitemShut {NoStop}%
\bibitem [{\citenamefont {Fukui}\ \emph {et~al.}(2005)\citenamefont {Fukui},
  \citenamefont {Hatsugai},\ and\ \citenamefont {Suzuki}}]{FHS05}%
  \BibitemOpen
  \bibfield  {author} {\bibinfo {author} {\bibfnamefont {T.}~\bibnamefont
  {Fukui}}, \bibinfo {author} {\bibfnamefont {Y.}~\bibnamefont {Hatsugai}},\
  and\ \bibinfo {author} {\bibfnamefont {H.}~\bibnamefont {Suzuki}},\
  }\bibfield  {title} {\bibinfo {title} {Chern numbers in discretized brillouin
  zone: Efficient method of computing (spin) hall conductances},\ }\bibfield
  {booktitle} {\emph {\bibinfo {booktitle} {Journal of the Physical Society of
  Japan}},\ }\href {https://doi.org/10.1143/JPSJ.74.1674} {\bibfield  {journal}
  {\bibinfo  {journal} {Journal of the Physical Society of Japan}\ }\textbf
  {\bibinfo {volume} {74}},\ \bibinfo {pages} {1674} (\bibinfo {year}
  {2005})}\BibitemShut {NoStop}%
\end{thebibliography}
%

\end{document}